\documentclass[pra,aps,amsmath,amssymb,amsfonts,twocolumn,nofootinbib,longbibliography,superscriptaddress]{revtex4-2}
\usepackage{amssymb}
\usepackage{bm,mathrsfs}
\usepackage{graphicx}
\usepackage{epsfig}
\usepackage{amsmath,bbm}
\usepackage{amsfonts,amssymb}
\usepackage{times}
\usepackage{verbatim}
\usepackage[sort&compress]{natbib}
\usepackage{amsmath}
\usepackage[colorlinks=true,citecolor=blue,linkcolor=blue,urlcolor=blue]{hyperref}
\usepackage[usenames]{color}

\newcommand{\bra}[1]{\langle #1|}
\newcommand{\ket}[1]{|#1\rangle}

\begin{document}

\title{Switchable selective interactions in a Dicke Model with Driven Biased term}

\author{Ning Yu}
\affiliation{Center for Quantum Sciences and School of Physics, Northeast Normal University, Changchun 130024, China}

\author{Shiran Wang}
\affiliation{Center for Quantum Sciences and School of Physics, Northeast Normal University, Changchun 130024, China}

\author{Chunfang Sun}
\affiliation{Center for Quantum Sciences and School of Physics, Northeast Normal University, Changchun 130024, China}

\author{Gangcheng Wang}
\email{wanggc887@nenu.edu.cn}
\affiliation{Center for Quantum Sciences and School of Physics, Northeast Normal University, Changchun 130024, China}

\date{\today}

\begin{abstract}
In this work, we propose a method to investigate controllable qubit-resonator interactions in a Dicke model with driven biased term. The nonlinearity of spectrum, which can be induced by qubit-resonator interactions, plays an important role in such controllable interactions. To gain insight into mechanism of the nonlinearity, we perform a unitary transformation to the Hamiltonian. The results show that the nonlinearity of the transformed Hamiltonian depends on the qubit-resonator coupling strength. The general forms of the effective Hamiltonians are discussed in detail based on the frequency modulation approach. The dynamical evolution can be switched on and off by adjusting the modulation parameters. By utilizing such controllable interactions, we discuss the creation of Dicke states and arbitrary superposition of Dicke states. We also consider the nonlinearity of energy level for the limit of large qubit numbers. In the thermodynamics limit, the kerr type nonlinearity is induced from ``magnon"-resonator coupling, and the selective preparation of ``magnon" Fock states can be studied under ``magnon" scenario.
\end{abstract}

\maketitle

\section{Introduction}
\label{SecI}
Quantum entanglement \cite{PRA2003Stockton,PRA2000Koashi,PRA2002Cabello,RMP2009Horodecki,book2010Nielsen}, as a non-classical nature of quantum mechanics, has been studied in various systems, including atoms \cite{PRL2002Hong,PRA2004Stockton,PRA2007Xiao,Epl2010Shao}, photons \cite{PRL2009Prevedel}, trapped ions \cite{PRL2009Wieczorek,PRL2012Noguchi,PRA2013Lamata}, and superconducting qubits \cite{PRL2009Fink,Nature2010Neeley,PRA2007Koch}. As a type of special multipartite entangled states, the Dicke states \cite{PR1954Dicke} have many important applications in multiparty quantum networking protocols \cite{PRL2009Prevedel}, quantum games \cite{NJPh2007Ozdemir}, quantum key distribution \cite{PRLett1991Ekert}, quantum memories \cite{NP2009Lvovsky}, efficient detection of inhomogeneous magnetic fields \cite{PRA2020Hakoshima}, quantum metrology \cite{PRA2012Toth} and so on. So far, various methods are used to create Dicke states such as adiabatic passage \cite{PRA2011Toyoda}, frequency-chirped optical pulses in a string of ions \cite{PRA2008Linington}, and Lyapunov control in circuit quantum electrodynamics (QED) systems \cite{RMP2021Blais,AOP2018Du}. In Ref. \cite{PRA2007Xiao}, the authors proposed that an $N$-qubit Dicke states can be prepared based on the so-called selective atom-cavity interaction in cavity QED systems.

On the other hand, periodic driving is an important tool to coherently manipulate the states of quantum systems \cite{ROPIP2017M,PR2004Chu,PRA2009Son,PRL2013Luo,RMP1990H,NJP2015Eckardt,AdP2015David}. The controllability of systems is largely extended when the periodic driving is introduced \cite{OE2020Wang}. Such versatile tool for quantum control is utilized to study many fields of quantum physics, such as exotic quantum properties of artificial quantum matter \cite{PRB2010Kitagawa}, Floquet analysis of a quantum system \cite{PRA2017Novi}, population trapping \cite{PRA2021Mallavarapu}, stochastic processes \cite{PR1993Peter}, and so on. By means of adjusting the parameters of periodic driving, one can realize various controllable selective interactions which arise from the possibility of adjusting parameters to resonance transition inside a chosen Hilbert subspace, while other transitions are suppressed \cite{MC2001Solano}.

In this work, we will study the biased Dicke model \cite{PR1954Dicke,JPB2013Daniel} with periodic driving. By means of time-periodic driving, the biased Dicke model can be modified. Various controllable interactions can be engineered by tuning the driving parameters. Then we can obtain different effective target Hamiltonians by utilizing the induced nonlinearity of the systems according to the initial and final states. Such special interactions are named selective resonance interactions \cite{PRA2017Wu}. By means of such special selective interactions, the Dicke states, as well as superposition Dicke states and other special entangled states can be generated coherently. In the limit of large $N$, the two-level atomic operators can be mapped to bosonic operators by means of Holstein-Primakoff transformation \cite{PR1940Holstein}. Then we can investigate nonlinearity and dynamics of the system under ``magnon" scenario.

The rest of this paper is organized as follows: In Sec. \ref{SecII}, we apply a suitable unitary transformation on the system. And we expand the Hamiltonian in the Dicke states and Fock states. Choosing the proper driving parameters, we can obtain a set of effective target Hamiltonians, such as Tavis-Cummings (TC) \cite{PR1968Tavis} and anti-TC interactions. In Sec. \ref{SecIII}, we discuss the applications of selective interactions, including Dicke states population trapping and the generations of entangled states such as Dicke states, Greenberger–Horne–Zeilinger (GHZ) states. In Sec. \ref{SecIV}, we consider the model in the thermodynamic limit case \cite{Sciences2018Zou}. The qubits collective operators can be replaced by the ``magnon" creation and annihilation operators based on the Holstein-Primakoff transformation. One can obtain the new general forms of the effective Hamiltonians based on the frequency modulation. By utilizing such controllable interactions, we discuss the creations of ``magnon" Fock states. Finally, we conclude this paper with summary and discussion in Sec. \ref{SecV}.

\section{The effective Hamiltonian for finite $N$}
\label{SecII}

We consider the system that is composed of $N$ biased qubits coupled to a single-mode bosonic field. To obtain the so-called switchable selective interactions, we apply a periodically driven field to the qubit system. Such frequency modulation will induce a series of qubit-resonator sidebands which will be applied to generate switchable selective interactions. The Hamiltonian that we consider in this work is given by (the Planck constant $\hbar = 1$ throughout the discussion)
\begin{equation}\label{eq_01}
  \hat{H}(t) = \hat{H}_r + \hat{H}_q(t) + \hat{H}_{\rm int},
\end{equation}
where
\begin{subequations}
  \begin{align}
     \hat{H}_r & = \omega_{r} \hat{a}^{\dagger} \hat{a},\nonumber\\
     \hat{H}_q(t) & = \sum_{m=1}^{N} \frac{\epsilon}{2} \hat{\sigma}_m^z + \sum_{m=1}^{N} \left[\frac{\Delta}{2} + \Omega_{d} \cos(\omega_{d}t ) \right]\hat{\sigma}_m^x,\nonumber \\
      \hat{H}_{\rm int} & = \sum_{m=1}^{N} g_{m} (\hat{a}^{\dagger} + \hat{a}) \hat{\sigma}_{m}^{x}.\nonumber
  \end{align}
\end{subequations}
Here $\hat{a}^{\dagger}(\hat{a})$ denote the creation (annihilation) operators of the bosonic field with frequency $\omega_r$, $\hat{\sigma}_{m}^{x} =\ket{e}_{m}\bra{g}+\ket{g}_{m}\bra{e}$, $\hat{\sigma}_{m}^{y} =-i \ket{e}_{m}\bra{g}+i\ket{g}_{m}\bra{e}$ and $\hat{\sigma}_{m}^{z}=\ket{e}_{m}\bra{e}-\ket{g}_{m}\bra{g}$ are the $m$-th qubit Pauli matrices, and $\ket{g}_{m}$ and $\ket{e}_{m}$ are the ground and excited states for the $m$-th qubit, respectively. Also, the notation $\epsilon$ is the frequency of the qubit and $\Delta$ is the energy split of the biased term. The notations $\Omega_d$ and $\omega_d$ are driving amplitude and frequency of the periodic driven field that will be used to achieve control of dynamics of the system. In the following, we assume the coupling strength between qubits and the bosonic mode is uniform (i.e. $g_{m}=g$). By introducing the qubits collective operators $\hat{J}_{\alpha}=\frac{1}{2}\sum_{m=1}^{N} \hat{\sigma}_{m}^{\alpha}$ ($\alpha=x,y,z$) and $\hat{J}_{\pm} = \hat{J}_{x} \pm i\hat{J}_{y}$. One can verify that the generators $\hat{J}_{\pm}$ and $\hat{J}_{z}$ satisfy the following su(2) relations: $[\hat{J}_{+}, \hat{J}_{-}]= 2 \hat{J}_{z}$ and $[\hat{J}_{z}, \hat{J}_{\pm}]= \pm \hat{J}_{\pm}$. Then the Hamiltonian in Eq. (\ref{eq_01}) can be rewritten in terms of generators as
\begin{equation}\label{eq_02}
\small
  \hat{H}(t) = \omega_{r} \hat{a}^{\dagger} \hat{a} + \epsilon \hat{J}_{z} + \left[ \Delta + 2\Omega_{d} \cos(\omega_{d}t) \right] \hat{J}_{x} +  2g (\hat{a}^{\dagger} + \hat{a})\hat{J}_{x}.
\end{equation}
In Ref. \cite{PRA2016Jaako}, the authors showed that strong nonlinearity of the energy spectrum can arise from ultrastrong qubit-resonator coupling. In order to get a clear insight, we perform the following unitary transformation to the system
\begin{equation*}\label{eq_02_1}
  \hat{R} =\exp \left[ -\alpha (\hat{a}^{\dagger} - \hat{a}) \hat{J}_{x} \right]\times \left[\prod_{m=1}^{N}\frac{1}{\sqrt{2}}(\hat{\sigma}_{m}^{x} + \hat{\sigma}_{m}^{z}) \right],
\end{equation*}
with $\alpha = 2g/\omega_{r}$, the transformed Hamiltonian is
\begin{equation}\label{eq_03}
  \hat{H}^{\prime}(t) = \hat{R}^{\dag} \hat{H}(t)\hat{R} = \hat{H}_{r}^{\prime}+\hat{H}_{q}^{\prime}(t)+ \hat{\tilde{H}}_{\rm int}^{\prime},
\end{equation}
where
\begin{subequations}\label{eq_04}
  \begin{align}
  \hat{H}_{r}^{\prime} & = \omega_{r}\hat{a}^{\dagger} \hat{a},\\
     \hat{H}_{q}^{\prime}(t) & = \left[\Delta+2\Omega_{d} \cos(\omega_{d}t)\right] \hat{J}_{z} - \frac{4g^2}{\omega_{r}} \hat{J}_{z}^2,\label{eq4b}\\
      \hat{H}_{\rm int}^{\prime} & = \frac{\epsilon}{2} \hat{D} (\alpha) \hat{J}_{+} +\frac{\epsilon}{2} \hat{D} (-\alpha) \hat{J}_{-}.
  \end{align}
\end{subequations}
Here $\hat{D} (\alpha) = \exp \left( \alpha \hat{a}^{\dagger} - \alpha^{*}\hat{a} \right)$ is displacement operator. The Eq. (\ref{eq_03}) shows that the qubit-qubit coupling is obtained by means of the unitary transformation. To show the properties of the spectrum without driving, we assume the driving amplitude $\Omega_{d}=0$. The eigenvalues of $\hat{H}_{0}^{\prime}(t)=\hat{H}_{r}^{\prime}+\hat{H}_{q}^{\prime}(t)$ are given as follows
\begin{equation}\label{eq_03_1}
  E_{n,k}=\Delta\left(k-\frac{N}{2}\right)-\frac{4g^{2}}{\omega_{r}}\left(k-\frac{N}{2}\right)^{2}+n\omega_{r},
\end{equation}
where $n$ and $k$ are the excitation number of the resonator and qubits. The degenerate for level $k$ is $C_{k}^{N}=N! /[k!(N-k)!]$. The Eq. (\ref{eq_03_1}) shows that energy levels depend on qubits excitation number $k$ nonlinearly, and its energy gap for different levels depends on the qubit-resonator coupling $g$ as follows
\begin{equation*}\label{eq_03_2}
  \Delta_{k}=E_{n,k+1}-E_{n,k}= \Delta + \frac{4 g^2}{\omega_r}(N-2k-1).
\end{equation*}

\begin{figure}
  \centering
  \includegraphics[angle=0,width=0.45\textwidth]{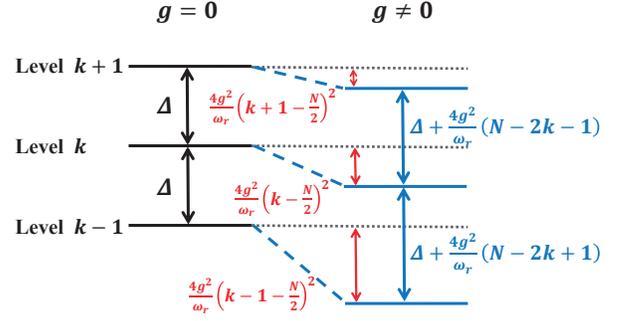}
  \caption{Energy spectrum for uncoupled case ($g = 0$) and coupled case ($g \neq 0$) from the Hamiltonian $\hat{H}_{q}^{\prime} = \Delta \hat{J}_{z} - \frac{4g^2}{\omega_{r}} \hat{J}_{z}^2$ with the driving amplitude $\Omega_{d}=0$ in Eq. (\ref{eq4b}). Because their energy gap depends on $k$, an appropriate drive field can be used to engineer selective interactions. }\label{fig1}
\end{figure}
\begin{figure}
  \centering
  \includegraphics[angle=0,width=0.45\textwidth]{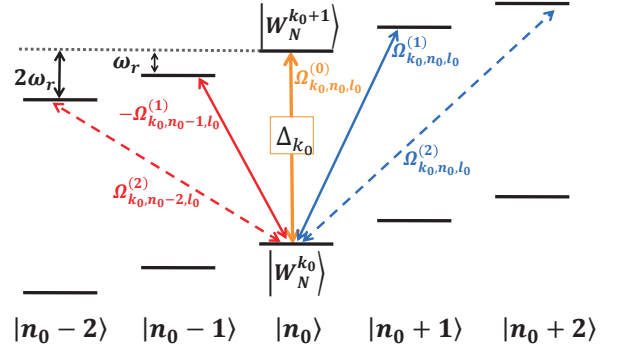}
  \caption{Energy level diagram with taking quantum state $\ket{W_{N}^{k_0}} \otimes \ket{n_0}$ as an example. Driving a carrier transition (solid yellow line) or $s_0$-th order blue sidebands (blue line) and red sidebands (red line) transition close to resonance results in an effective interaction.}\label{fig3}
\end{figure}

The nonlinear energy levels of the $N$-qubit systems without driving are shown in Fig. \ref{fig1}. It can be clearly seen that the energy level separation of $N$-qubit systems is equal with the coupling strength $g = 0$. But separation in energy level is not equal with the coupling strength $g \neq 0$, and depends on the excitation number of the qubits. Such properties of energy level can be utilized to generate $N$-qubit Dicke states, as well as superposition Dicke states. To achieve the full control of the systems, we consider the driving amplitude $\Omega_{d}\neq 0$. Then we can achieve so-called controlled selective transitions by adjusting the driving parameters.
Moving to the rotating framework defined by
\begin{equation}\label{u}
  \begin{split}
     \hat{U}_{0}(t) &= \exp \left[ -i \int_{0}^{t} dt^{\prime} \hat{H}_{0}^{\prime}(t^{\prime})  \right] \\
       & =\exp \left[ -it\omega_{r}\hat{a}^{\dagger} \hat{a} -i\hat{F}\left(\hat{J}_{z}\right)\right],
  \end{split}
\end{equation}
where
\begin{equation*}
  \hat{F}\left(\hat{J}_{z}\right)=[\Delta t + \eta_{d}\sin(
\omega_{d}t)]\hat{J}_{z}-\frac{4g^2t}{\omega_{r}}\hat{J}_{z}^{2},
\end{equation*}
with $\eta_{d}=2\Omega_{d}/\omega_{d}$, one obtains the following transformed Hamiltonian \cite{PRA1993Kitagawa}
\begin{equation}\label{HI}
   \begin{split}
      \hat{H}_{I}^{\prime}(t) =& \hat{U}_{0}^{\dagger}(t)\left[ \hat{H}^{\prime}(t)-i\partial_{t} \right]\hat{U}_{0}(t) , \\
       =&\frac{\epsilon}{2} \hat{D} (\alpha e^{i \omega_{r}t}) \hat{J}_{+} e^{i \hat{f}\left(\hat{J}_{z}\right)}+ {\rm H.c.},
   \end{split}
\end{equation}
where
\begin{equation*}
   \begin{split}
      \hat{f}\left(\hat{J}_{z}\right) &= \hat{F}\left(\hat{J}_{z}+1\right)-\hat{F}\left(\hat{J}_{z}\right),\\
      &=\left[\Delta-\frac{4g^2}{\omega_{r}}\left(2\hat{J}_{z}+1\right) \right] t+ \eta_{d}\sin(\omega_{d}t) .
         \end{split}
\end{equation*}
Using the Jacobi-Anger identity
\begin{equation}\label{J}
  \exp( i z \sin x) = \sum_{l=-\infty}^{\infty} \mathcal{J}_{l} \left(z \right) e^{i l x },
\end{equation}
with $\mathcal{J}_{l}(z)$ is the $l$-th order Bessel function of the first kind, the Hamiltonian (\ref{HI}) can be rewritten
\begin{equation}
\begin{split}
      \hat{H}_{I}^{\prime}(t)
       =&\frac{\epsilon}{2} \hat{D} (\alpha e^{i \omega_{r}t}) \hat{J}_{+} \sum_{l=-\infty}^{\infty} \mathcal{J}_{l}(\eta_d)  e^{i \left[l \omega_{d} + \Delta -\frac{4g^2}{\omega_{r}} \left( {2\hat{J}_{z}} + 1 \right) \right] t}\nonumber\\
       &+ {\rm H.c.}.
\end{split}
\end{equation}
To obtain a closed analytical description of Dicke states and superposition Dicke states generation, we introduce the normalized $N$-qubit Dicke states with $k$ atomic excitations \cite{PRL2012Noguchi,PRA2011Zhou,PRA2009Hume}
\begin{equation*}
    \ket{W_{N}^{k}} \equiv \left( C_{k}^{N} \right)^{ -1 / 2} \sum_{m=1}^{C_{k}^{N}} P_{m} \ket{e_{j_{1}}, e_{j_{2}}, \cdots, e_{j_{k}}, g_{j_{k+1}} \cdots, g_{j_{N}}}.
\end{equation*}
Here $\sum_{m=1}^{C_{k}^{N}} P_{m}(\bullet)$ indicates the sum over all particle permutations. In the Dicke state basis, the collective operators can be reduced to $\hat{J}_{z, \pm}^{W}$ as follows
\begin{subequations}\label{2.9}
  \begin{align}
        \hat{J}_{z}^{W} &= \sum_{k=0}^{N} \left(k-\frac{ N}{2}\right) \ket{W_{N}^{k}} \bra{W_{N}^{k}}, \\
        \hat{J}_{+}^{W} &= \sum_{k=0}^{N} h(k) \ket{W_{N}^{k+1}} \bra{W_{N}^{k}}, \\
        \hat{J}_{-}^{W} &= \sum_{k=0}^{N} h(k) \ket{W_{N}^{k}} \bra{W_{N}^{k+1}},
    \end{align}
\end{subequations}
where $h(k) = \sqrt{(k+1)(N-k)}$. In the Dicke state basis, the system reduce to a spin-$N/2$ system, and the Hilbert space reduces from $2^{N}$ to $(N+1)$. The displacement operator can be rewritten in the Fock state basis (cf. Appendix \ref{appendix1}),
\begin{equation}\label{eq_06}
     \hat{D}(\alpha e^{i\omega_r t}) = \sum_{m,n=0}^{\infty} \bra{m}\hat{D}(\alpha e^{i\omega_r t})\ket{n} \hat{A}_{m,n},
\end{equation}
where $\hat{A}_{m, n} = \ket{m} \bra{n}$ and the matrix elements $D_{mn} =\bra{m}\hat{D}(\alpha e^{i\omega_r t})\ket{n}$ read
\begin{equation*}\label{eq_07}
D_{mn}=\left\{\begin{array}{ll}
e^{-\frac{1}{2}\alpha^{2}}  L_{n}^{(0)}\left(\alpha^{2}\right), & m = n; \\
e^{-\frac{1}{2}\alpha^{2}} \alpha^{s} e^{is\omega_r t}\sqrt{\frac{n !}{m !}} L_{n}^{(s)}\left(\alpha^{2}\right), & m > n;\\
e^{-\frac{1}{2}\alpha^{2}} (-\alpha)^{s}e^{-is\omega_r t} \sqrt{\frac{m !}{n !}} L_{m}^{(s)}\left(\alpha^{2}\right), & m<n.
\end{array}\right.
\end{equation*}
Here $L_{m}^{(s)}(\alpha^{2})$ is an associated Laguerre polynomial with $s=|m-n|$. The cases $s=0$ and $s > 0$ correspond to carrier transition and the $s$-th sidebands transition, respectively. In terms of Dicke states and Fock states, the Hamiltonian Eq. (\ref{HI}) can be recast as follows
\begin{equation*}\label{eq_11}
    \begin{split}
  \hat{H}_{I}(t) =& \sum_{nkl}\sum_{s=1}^{\infty} \left[\Omega_{knl}^{(0)}(\alpha,\eta_d) e^{i \delta^{(0)}_{kl} t} \hat{W}_{k+1,k} \otimes \hat{A}_{n,n} \right. \\
                 &+ (-1)^s \Omega_{knl}^{(s)}(\alpha,\eta_d) e^{i \delta^{(-s)}_{kl} t} \hat{W}_{k+1,k} \otimes \hat{A}_{n,n+s} \\
                 &+ \left. \Omega_{knl}^{(s)}(\alpha,\eta_d) e^{i \delta^{(+s)}_{kl} t} \hat{W}_{k+1,k} \otimes \hat{A}_{n+s,n}  \right]+
                 \rm H.c.,
     \end{split}
\end{equation*}
where $\hat{W}_{k,k^{\prime}} = \ket{W_{N}^{k}} \bra{W_{N}^{k^{\prime}}}$ and
\begin{subequations}
    \begin{align}
          \Omega_{knl}^{(s)}&= \frac{\epsilon }{2}h(k)\mathcal{J}_{l}(\eta_d) \sqrt { \frac{n!}{(n+s)!}}  e^{ - \frac{1}{2}\alpha^{2}}  \alpha^{s} L_{n}^{(s)} (\alpha^{ 2 }) ,\nonumber\\
          \delta_{kl}^{(\pm s)} &= l\omega_d + \Delta_{k} \pm s\omega_{r}.\nonumber
    \end{align}
\end{subequations}
Obviously, we can tune the driving parameters to obtain different effective Hamiltonians. By tuning the driving frequency $\omega_d$, one can obtain the desired selective interactions based on rotating-wave approximation (RWA). It is worth mentioning that the strengths of effective interactions can be tuned by adjusting the ratio $\eta_d$ \cite{PRA2015Xue}. Such properties can be utilized to switch on and off the interactions coherently.
As shown as Fig. \ref{fig3}, taking quantum state $\ket{W_{N}^{k_0}} \otimes \ket{n_0}$ as an example, we plot the energy level diagram with considering the carrier (solid yellow line), the $s_0$-th order red sidebands (red lines), and blue sidebands (blue lines). Driving a carrier transition or $s_0$-th order sideband transition close to resonance results in an effective interaction with coupling constant. Then the effective Hamiltonians can be obtained approximately.

When we tune the driving frequency $\omega_d = -\Delta_{k_0}/l_{0}$, only the terms containing the operator combinations of the kind $\hat{W}_{k_0+1,k_0} \otimes \hat{A}_{n_0,n_0}$  and $\hat{W}_{k_0,k_0+1} \otimes \hat{A}_{n_0,n_0}$  are time-independent. One also can verify that the following conditions $\left| \delta_{kl}^{(0)} \right| \gg \left|\Omega_{knl}^{(0)}(\alpha,\eta_d) \right|$ for $k \neq k_0$ or $l \neq l_0$, $\left| \delta_{kl}^{(+s)} \right| \gg \left|\Omega_{knl}^{(s)}(\alpha,\eta_d) \right|$ and $\left| \delta_{kl}^{(-s)} \right| \gg \left|\Omega_{knl}^{(s)}(\alpha,\eta_d) \right|$ are satisfied in the ultrastrong coupling regime. Let the initial state be $\ket{W_{N}^{k_0}} \otimes \ket{n_0}$. Only the transition $\ket{W_{N}^{k_0}} \otimes \ket{n_0} \leftrightarrow \ket{W_{N}^{k_0+1}} \otimes \ket{n_0}$ (solid yellow line in Fig. \ref{fig3}) is permitted and the other transitions are suppressed. 
The system evolution is dominated by the following effective Hamiltonian
\begin{equation}\label{eq_12}
\small
    \hat{H}_{\rm eff}^{\rm car} = \Omega_{k_0,n_0,l_0}^{(0)}(\alpha,\eta_d) \left( \hat{W}_{k_0+1,k_0} \otimes \hat{A}_{n_0,n_0} + \rm H.c.\right).
\end{equation}
When driving frequency $\omega_d = -(\Delta_{k_0}-s_0 \omega_r)/l_0$ is tuned, and the conditions $\left| \delta_{kl}^{(-s)} \right| \gg \left| \Omega_{knl}^{(s)}(\alpha,\eta_d) \right|$ for $k \neq k_0$ or $s\neq s_0$ or $l \neq l_0$, $\left| \delta_{kl}^{(+s)} \right| \gg \left| \Omega_{knl}^{(s)}(\alpha,\eta_d) \right|$ and $\left| \delta_{kl}^{(0)} \right| \gg \left| \Omega_{knl}^{(0)}(\alpha,\eta_d) \right|$ are satisfied in the ultrastrong coupling regime. The transition $\ket{W_{N}^{k_0}} \otimes \ket{n_0+s_0} \leftrightarrow \ket{W_{N}^{k_0+1}} \otimes \ket{n_0}$ (solid red line in Fig. \ref{fig3}) is allowed when the initial state is prepared on $\ket{W_{N}^{k_0}} \otimes \ket{n_0+s_0}$. The corresponding effective TC Hamiltonian for such dynamical evolution reads
\begin{equation}\label{eq_13}
\small
    \hat{H}_{\rm eff}^{\rm TC}  = (-1)^{s_0} \Omega_{k_0,n_0,l_0}^{(s_0)}(\alpha,\eta_d) \left(\hat{W}_{k_0+1,k_0} \otimes \hat{A}_{n_0,n_0+s_0} + \rm H.c.\right).
\end{equation}
When the driving frequency $\omega_d = -(\Delta_{k_0} + s_0 \omega_r)/l_0$ is tuned, and the conditions $\left| \delta_{kl}^{(+s)} \right| \gg \left|\Omega_{knl}^{(s)}(\alpha,\eta_d) \right|$ for $k \neq k_0$ or $s\neq s_0$ or $l \neq l_0$, $\left| \delta_{kl}^{(-s)} \right| \gg \left| \Omega_{knl}^{(s)}(\alpha,\eta_d) \right|$ and $\left| \delta_{kl}^{(0)} \right| \gg \left| \Omega_{knl}^{(0)}(\alpha,\eta_d) \right|$ are satisfied in the ultrastrong coupling regime. Only the transition $\ket{W_{N}^{k_0}} \otimes \ket{n_0} \leftrightarrow \ket{W_{N}^{k_0+1}} \otimes \ket{n_0+s_0}$ (solid blue line in Fig. \ref{fig3}) is allowed when the initial state is prepared on $\ket{W_{N}^{k_0}} \otimes \ket{n_0}$. Accordingly, the effective anti-TC Hamiltonian reads
\begin{equation}\label{eq_14}
\small
    \hat{H}_{\rm eff}^{\rm aTC} = \Omega_{k_0,n_0,l_0}^{(s_0)}(\alpha,\eta_d) \left(\hat{W}_{k_0+1,k_0} \otimes \hat{A}_{n_0+s_0,n_0} + \rm H.c.\right).
\end{equation}
Then we obtain the so-called effective Hamiltonians for selective interactions by tuning the driving parameters. In the following sections, we will study the applications of the effective Hamiltonians.

\section{The applications of selective interaction}
\label{SecIII}

\subsection{Dicke states population trapping}
\label{SecIIIA}

\begin{figure}
  \centering
  \includegraphics[angle=0,width=0.48\textwidth]{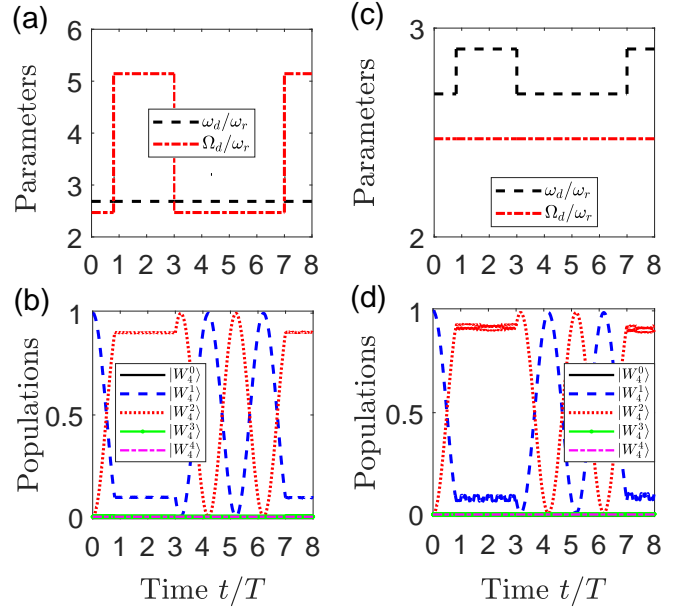}
  \caption{(a) and (c) show the driving amplitude $\Omega_d$ and frequency $\omega_d$ versus time $t$. (b) and (d) show the populations of Dicke states versus time $t$, such as the black line $P_{\ket{W_{4}^{0}}} = |\langle \tilde{\psi}(t)| W_{4}^{0},0\rangle|^{2}$, where $\ket{\tilde{\psi}(t)}$ is the evolution state of the full Hamiltonian (\ref{eq_03}) with the initial state $\ket{\psi(0)}=\ket{W_{4}^{1}}\otimes\ket{0}$.
  The parameters are chosen as $\epsilon=0.01\omega_{r}$, $\Delta=2\pi \times 5.4$ GHz, $g=0.24\omega_{r}$ and $\omega_{r}=2\pi \times 2.2$ GHz. Time $t$ displays as multiples of $T = \pi/[2\Omega_{1,0,1}^{(0)}] \approx 1.7893 \times 10^{-8}$s.}\label{fig_01}
\end{figure}

Population trapping \cite{PRA2021Mallavarapu}, a phenomenon named dynamical stabilization, plays a key role in both classical and quantum mechanical systems. Periodically tuning the system parameters in time, dynamically unstable configuration of a system can be stabilized. The Rabi coupling strength depending on the periodic modulation amplitude and frequency can be suppressed, leading to dynamical stabilization of the initial state.

According to the effective Hamiltonians in Eqs. (\ref{eq_12}-\ref{eq_14}), the effective Rabi frequencies $\Omega_{knl}^{(s)}(\alpha,\eta_{d})$ are proportional to the Bessel function $\mathcal{J}_{l}(\eta_d)$. One can tune the ratio of driving amplitude and frequency to switch on/off the interactions. When the ratio $\eta_{d}=2\Omega_d/\omega_d$ is set to the zeros of Bessel function $\mathcal{J}_{l}(\eta_{d})$ (i.e. $\mathcal{J}_{l}(\eta_{d})=0$), the dynamical evolution is switched off and the corresponding transition is strongly suppressed. Such phenomenon is named population trapping. We will show how to switch on/off the interactions by tuning the driving amplitude with an example. If we tune the driving frequency $\omega_{d}=\Delta_{k_{0}}$ (i.e. $l_{0}=-1$) and nonzero $\mathcal{J}_{-1}(\eta_{d})$, the interaction in Eq. (\ref{eq_12}) is switched on. Let $\ket{\psi(0)}=\ket{W_{4}^{1}}\otimes\ket{0}$ (i.e. $k_{0}=1$, $n_{0}=0$) be the initial state. The Fig. \ref{fig_01}(b) shows the switchable dynamical evolution for the case $N=4$ when the driving parameters are tuned as the Fig. \ref{fig_01} (a). First, we tune the driving frequency $\omega_{d}=\Delta_{1} \approx 2.6849\omega_r$ and amplitude $\Omega_{d} = 0.92\omega_d \approx 2.4701 \omega_r $ as shown in Fig. \ref{fig_01} (a). In this case, $\mathcal{J}_{-1}(\eta_{d}) \approx -0.582$, the evolution state approximatively reads
\begin{equation}\label{eq_15}
\small
     \ket{\psi_{1}(t)}= \left[\cos \left(\Omega_{1,0,1}^{(0)}t \right)\ket{W_{4}^{1}} + i\sin\left(\Omega_{1,0,1}^{(0)}t \right)\ket{W_{4}^{2}}\right]\otimes\ket{0}.
\end{equation}
Here we used the relationship $\mathcal{J}_{-l}(\eta_d)=(-1)^{l}\mathcal{J}_{l}(\eta_d)$. The probabilities of the qubits being in the initial state $\ket{\psi(0)}$ and the final state $\ket{\psi_{f}}=\ket{W_{4}^{2}}\otimes\ket{0}$ at time $t$ are then given by $\left|\cos (\Omega_{1,0,1}^{(0)}t )\right|^{2}$ and $\left|\sin (\Omega_{1,0,1}^{(0)}t )\right|^{2}$, correspondingly.
The inversion is given by
\begin{equation}\label{3.6}
     \small
     Q(t)\equiv \left|\cos \left(\Omega_{1,0,1}^{(0)}t \right)\right|^{2}-\left|\sin \left(\Omega_{1,0,1}^{(0)}t \right)\right|^{2}=\cos \left(\Omega_{R}t \right).
\end{equation}
The inversion oscillates between -1 and 1 at Rabi frequency $\Omega_{R}=2\Omega_{1,0,1}^{(0)}(\alpha,\eta_d)$. Correspondingly, the half period of dynamic evolution (Rabi oscillation) is $T=\pi/\Omega_{R}\approx 1.7893 \times 10^{-8}$s.
At time $t_1=4T/5 \approx 1.4314 \times 10^{-8}$s, we tune the amplitude $\Omega_{d} \approx 5.1439\omega_r$. In this case, $\mathcal{J}_{-1}(\eta_{d}) \approx 0$, and the interaction is switched off. Then the state $\ket{\psi_{1}(t_{1})}$ is prepared and storaged as folllows
\begin{equation}\label{psi}
\small
  \ket{\psi_{1}(t_{1})}= \left[\cos \left(\frac{2\pi}{5}\right)\ket{W_{4}^{1}} +i\sin\left(\frac{2\pi}{5}\right)\ket{W_{4}^{2}}\right]\otimes\ket{0}.
\end{equation}
At time $t_2= 3T \approx 5.3679 \times 10^{-8}$s, we tune the amplitude $\Omega_{d} = 2.4701 \omega_r$. Then the interaction is switched on and the evolution state is governed by the Hamiltonian in Eq. (\ref{eq_12}). At time $t_3=7T \approx 1.2525 \times 10^{-7}$s, we tune the amplitude $\Omega_{d} = 5.1439\omega_r$ again, and the interaction is switched off accordingly.

In fact, one also can tune the driving frequency to realize Dicke states population trapping. When the frequency $ \omega_d \neq -\Delta_{k_0}/l_0 $ and $\omega_d \neq -\left(\Delta_{k_0} \pm s_0 \omega_r \right)/l_0 $, the dynamical evolution is switched off and the corresponding transition is forbidden. The Fig. \ref{fig_01}(d) shows the switchable dynamical evolution which is similar to Fig. \ref{fig_01}(b) when the driving parameters are chosen as Fig. \ref{fig_01}(c). First, we tune the driving parameters $\omega_{d}=\Delta_{1} \approx 2.6849\omega_r$ and $\Omega_{d} = 0.92\omega_d \approx 2.4701 \omega_r $ as shown in Fig. \ref{fig_01}(c), which are consistent with Fig. \ref{fig_01}(a) at the initial time, and the evolution states are shown as Eq. (\ref{eq_15}). At time $t_1=4T/5$, we tune the frequency $\omega_{d} \approx 2.90\omega_r$. In this case, $\omega_d \in (\Delta_{0},\Delta_{1})$, none of the effective resonance frequencies is satisfied in Eqs. (\ref{eq_12}-\ref{eq_14}), and the interaction is switched off. Then the state $\ket{\psi_{1}(t_{1})}$ is prepared and storaged. At time $t_2= 3T$, we tune the frequency $\omega_{d} = \Delta_{1} \approx 2.6849\omega_r$ again. Then the interaction is switched on and the evolution state is governed by the Hamiltonian in Eq. (\ref{eq_12}). At time $t_3=7T$, we tune the frequency $\omega_{d} = 2.90\omega_r$ again, and the interaction is switched off accordingly. The other parameters are chosen as the same as in Fig. \ref{fig_01}.

\subsection{The generation of Dicke states}
\label{SecIIIB}
\begin{figure}[htbp]
    \centering
    \includegraphics[angle=0,width=0.48\textwidth]{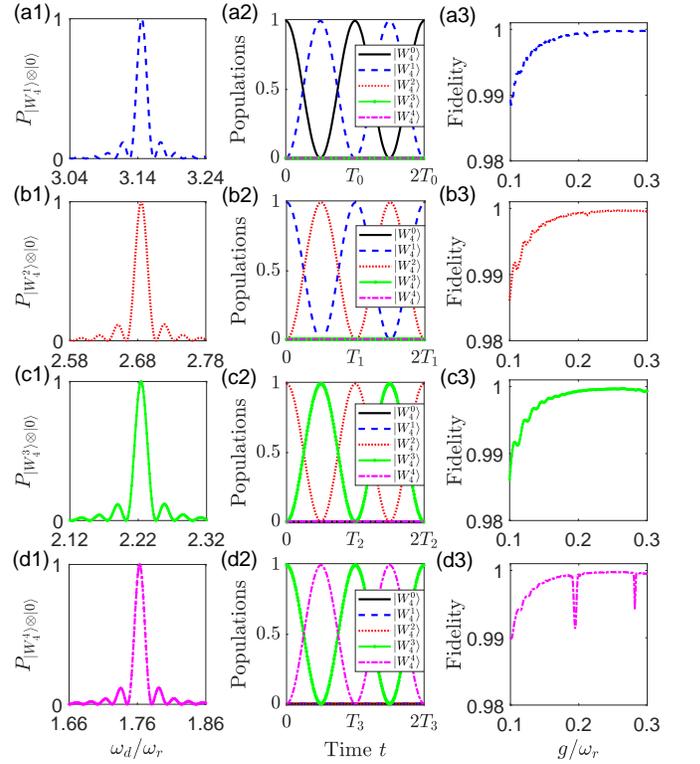}
    \caption{Numerical results of a four-qubit system based on the selective resonant interaction, where the parameters are chosen as $\epsilon=0.01\omega_{r}$, $\Delta=2\pi \times 5.4$ GHz, and $\omega_{r}=2\pi \times 2.2$ GHz. (a1)-(d1) show the population of the states $\ket{W_4^1}$, $\ket{W_4^2}$, $\ket{W_4^3}$, and $\ket{W_4^4}$ generated from $\ket{W_4^0}$, $\ket{W_4^1}$, $\ket{W_4^2}$, and $\ket{W_4^3}$ versus $\omega_d/\omega_r$, respectively. (a2)-(d2) show the populations of the states $\ket{W_4^0}$, $\ket{W_4^1}$, $\ket{W_4^2}$, $\ket{W_4^3}$ and $\ket{W_4^4}$ versus time $t$ with the couple strength $g=0.24\omega_{r}$. Time $t$ displays as multiples of $T_0=\pi/[\Omega_{0,0,1}^{(0)}]$, $T_1=\pi/[\Omega_{1,0,1}^{(0)}]$, $T_2=\pi/[\Omega_{2,0,1}^{(0)}]$, and $T_3=\pi/[\Omega_{3,0,1}^{(0)}]$. (a3)-(d3) show the fidelities of the states $\ket{W_4^0} \to \ket{W_4^1}$, $\ket{W_4^1} \to \ket{W_4^2}$, $\ket{W_4^2} \to \ket{W_4^3}$, and $\ket{W_4^3} \to \ket{W_4^4}$ versus $g/\omega_r$, such as $\mathcal{F}_{\ket{W_4^1}} = |\langle \tilde{\psi}(t)| \psi(t) \rangle|^{2}$, where $\ket{\tilde{\psi}(t)}$ and $\ket{\psi(t)}$ are the evolution states of the full Hamiltonian and the effective Hamiltonian at time $t=T_0/2$, respectively.}
    \label{fig_03}
\end{figure}

In this section, we will show how to generate the Dicke states with the carrier transition [i.e. Eq. (\ref{eq_12})]. Let $\ket{\psi_0}=\ket{W_{4}^{0}} \otimes \ket{0}$ (i.e. $k_0=0$, $n_0=0$) be the initial state, and $\ket{W_{4}^{1}} \otimes \ket{0}$ be the target state. Considering the initial and final states, one can adjust the driving frequency to realize the selective interaction, because of varying the detuning in time. The peak of population of state $\ket{W_{4}^{1}} \otimes \ket{0}$ from the initial state appears at $\omega_d/\omega_r \approx 3.1457$ corresponding to $\omega_d = \Delta_{0}$ (i.e. $l_0=0$) in Fig. \ref{fig_03}(a1), and we plot the variations of the states populations in $\ket{W_{4}^{0}}$, $\ket{W_{4}^{1}}$, $\ket{W_{4}^{2}}$, $\ket{W_{4}^{3}}$ and $\ket{W_{4}^{4}}$ with time $t$ which displays as multiples of $T_0=\pi/[\Omega_{0,0,1}^{(0)}] \approx 4.3828 \times 10^{-8}$s, as shown in Fig. \ref{fig_03}(a2). We find nearly perfect oscillations between the states $\ket{W_{4}^{0}}$ and $\ket{W_{4}^{1}}$, and almost zero population in the other states when the driving frequency is selected $\omega_d = \Delta_{0}$ and the other parameters are shown as the Fig. \ref{fig_03}. The evolution states are shown as follows
\begin{equation}\label{eq_16}
\small
     \ket{\psi_{2}(t)}= \left[\cos \left(\Omega_{0,0,1}^{(0)}t \right)\ket{W_{4}^{0}} + i\sin\left(\Omega_{0,0,1}^{(0)}t \right)\ket{W_{4}^{1}}\right]\otimes\ket{0}.
\end{equation}
Therefore, one can tune the $\Omega_d$ or $\omega_d$ to switch off the interaction at time $t_0$ as described in Sec. \ref{SecIIIA}, the target Dicke state $\ket{W_4^1}$ or superposition Dicke states [cf. Eq. (\ref{eq_16})] of $\ket{W_{4}^{0}}$ and $\ket{W_4^1}$ can be generated and stored. It is clearly shown in Fig. \ref{fig_03}(a3) that the fidelity $\mathcal{F}= |\langle \tilde{\psi}(T_0/2)| \psi_{2}(T_0/2)\rangle|^2$ \cite{JMO1994Richard} increases with increasing value of the ratio $g/\omega_r$, where $\ket{\tilde{\psi}(T_0/2)}$ is the evolution state controlled by the full Hamiltonian Eq. (\ref{eq_03}) with the initial state $\ket{\psi(0)}$ at time $t=T_0/2$. The following parameters are shown in Fig. \ref{fig_03}. We can see that when $g/\omega_r = 0.1$, the fidelity $\mathcal{F} \approx 0.9893$, and when $g/\omega_r = 0.3$, the fidelity $\mathcal{F} \approx 0.9998$. The reason is that when the value of the coupling strength $g$ between qubits and the bosonic mode is increasing, the nonlinear of energy spectrum is increasing accordingly. Our scheme relies on the nonlinearity of energy spectrum, and hence different choices of $g$ give different fidelities of the target state. That is also why we do not consider the case of Lamb-Dicke approximation. In Fig. \ref{fig_03}(b)-(d), the other Dicke states $\ket{W_4^2}$, $\ket{W_4^3}$, $\ket{W_4^4}$ also can be generated with high fidelity, when the initial state is given. And in Tab. \ref{tab-III}, we list the fidelities of Dicke states $\ket{W_{N}^{k}}$ from the initial state $\ket{W_{N}^{0}}$ with $g=0.2895\omega_{r}$.
\begin{table}[h]
\centering
\caption{The fidelities of Dicke states for different $N$ with the initial state $\ket{W_{N}^{0}}$. The parameters are chosen as $\epsilon=0.01\omega_{r}$, $\Delta=2\pi \times 5.4$ GHz, $g=0.2895\omega_{r}$ and $\omega_{r}=2\pi \times 2.2$ GHz.}
\label{tab-III}
\begin{tabular}{cccccccc}
\hline
\hline
  $N$ & $\mathcal{F}_{\ket{W_{N}^{1}}}$ & $\mathcal{F}_{\ket{W_{N}^{2}}}$ & $\mathcal{F}_{\ket{W_{N}^{3}}}$ & $\mathcal{F}_{\ket{W_{N}^{4}}}$ & $\mathcal{F}_{\ket{W_{N}^{5}}}$ & $\mathcal{F}_{\ket{W_{N}^{6}}}$ & $\mathcal{F}_{\ket{W_{N}^{7}}}$\\
\hline
   3 & 0.9999 & 0.9995 & 0.9995 &  &  &  &\\
   4 & 0.9998 & 0.9995 & 0.9991 & 0.9989 &  &  &\\
   5 & 0.9997 & 0.9992 & 0.9983 & 0.9974 & 0.9964 &  &\\
   6 & 0.9996 & 0.9986 & 0.9979 & 0.9963 & 0.9946 & 0.9901 & \\
   7 & 0.9995 & 0.9984 & 0.9970 & 0.9955 & 0.9938 & 0.9916 & 0.9669\\
\hline
\hline
\end{tabular}
\end{table}

\subsection{The generation of superposition Dicke states}
\label{SecIIIB}

\begin{figure}
    \centering
    \includegraphics[angle=0,width=0.48\textwidth]{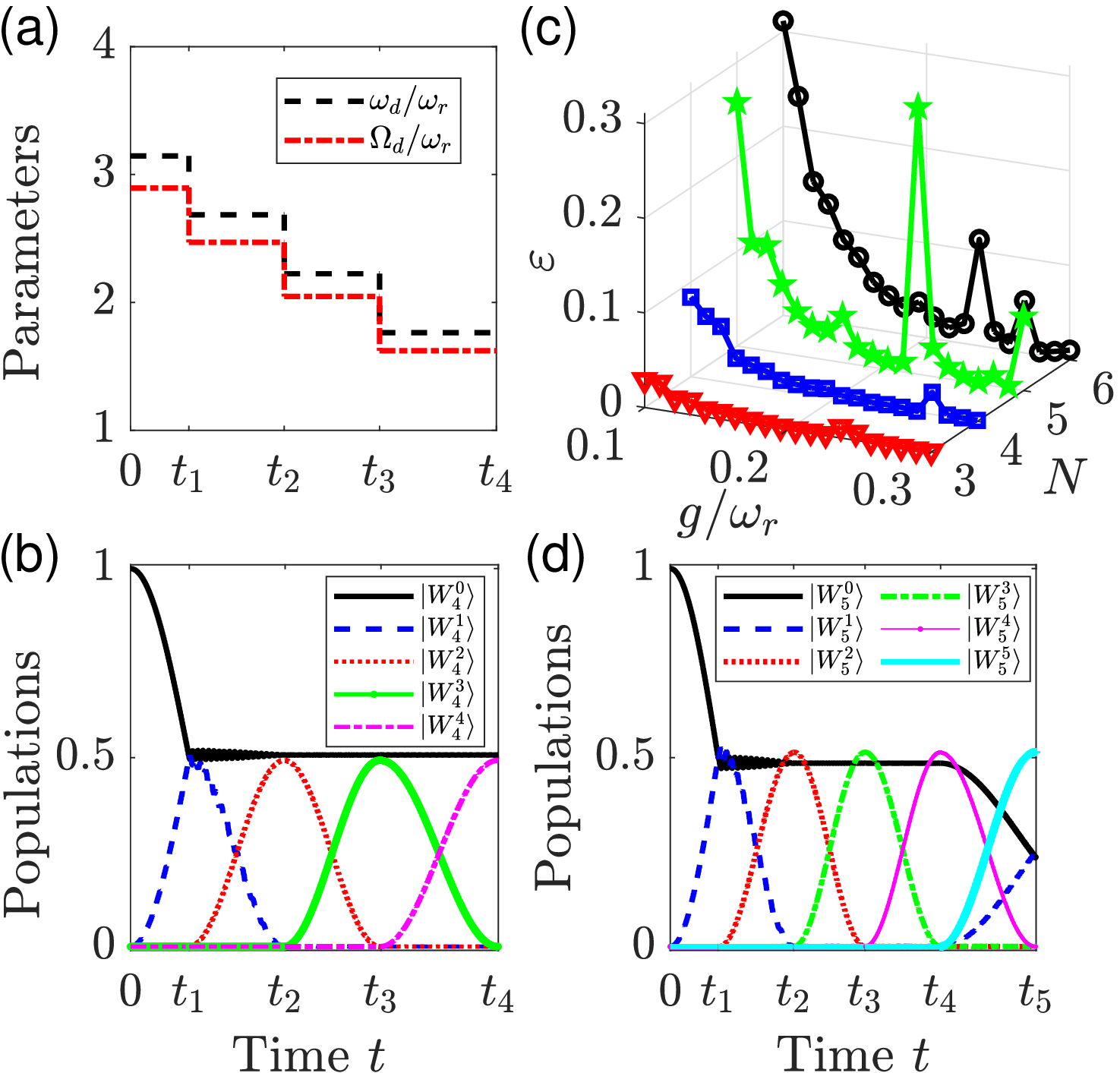}
    \caption{The parameters are chosen as $\epsilon=0.01\omega_{r}$, $\Delta=2\pi \times 5.4$ GHz, and $\omega_{r}=2\pi \times 2.2$ GHz. (a) shows the driving amplitude $\Omega_d$ and frequency $\omega_d$ versus time $t$. (b) shows the populations of Dicke states from $\ket{W_{4}^{0}}$ versus time $t$ with $g=0.24\omega_{r}$. Time $t_1=\pi/[4\Omega_{0,0,1}^{(0)}]$, $t_2 = t_1 + \pi/[2\Omega_{1,0,1}^{(0)}]$, $t_3 = t_2 + \pi/[2\Omega_{2,0,1}^{(0)}]$, and $t_4 = t_3 + \pi/[2\Omega_{3,0,1}^{(0)}]$. (c) shows the fidelity error $\varepsilon=1-\mathcal{F}$ of GHZ state for different $N$-qubits versus the coupling strength $g/\omega_r$. The red triangle line is $N=3$, blue quadrilateral line is $N=4$, green star line is $N=5$, and black circle line is $N=6$. (d) shows the populations of Dicke states for $N=5$ from state $\ket{W_{5}^{0}}$ versus time $t$ with $g = 0.2263\omega_r$.}
    \label{fig_04}
\end{figure}

With the controllable interaction, we can generate arbitrary superposition of Dicke states. To show how to generate superposition Dicke states, we take the generation of GHZ state for an example. The GHZ state which was first studied by Daniel Greenberger, Michael Horne and Anton Zeilinger in 1989, is one of the superposition Dicke states \cite{AmJPh1990Greenberger}. It can be used in protocols of quantum communication and cryptography, for example, in secret sharing \cite{PRLett1991Ekert,QC2014Bennett}.
The $N$-qubit GHZ state reads
\begin{equation}
\begin{split}
  \ket{GHZ}_{N} &= \frac{1}{\sqrt{2}}\left( \ket{g_{1} g_{2} \cdots g_{N}} + e^{i \varphi} \ket{e_{1} e_{2} \cdots e_{N}} \right)\\
  &= \frac{1}{\sqrt{2}} \left( \ket{W_{N}^{0}} + e^{i \varphi} \ket{W_{N}^{N}} \right).
\end{split}
\end{equation}
As shown in Fig. \ref{fig_04}(a) and Fig. \ref{fig_04}(b), we plot the creation of four-qubit GHZ state (i.e. $N=4$) with the initial state $\ket{\psi(0)}=\ket{W_{4}^{0}}\otimes\ket{0}$ (i.e. $k_0=1$, $n_0=0$). One can implement different controllable selective carrier resonance interactions in Eq. (\ref{eq_12}) by tuning the driving parameters $\omega_d$ and $\Omega_d$, where the other parameters are chosen as the Fig. \ref{fig_04}.
The driving frequencies $\omega_{d}$ of different resonance conditions are shown in Fig. \ref{fig_03}(a1)-(d1) and the quantum evolution states corresponding to different special moments $t$ are shown in the Tab. \ref{tab-I}. It is clear that the GHZ state is generated at time $t_4$. The fidelity of the state $\ket{\psi(t_4)}$ is $\mathcal{F} \approx 0.998$ by numerical simulation.

Obviously, the overall trend of the fidelity error $\varepsilon=1-\mathcal{F}$ for the GHZ state increases with the number of the qubits $N$ in Fig. \ref{fig_04}(c). In Tab. \ref{tab-II}, we list the fidelities of GHZ states from the initial excitation number $k_{0}=0$ with $g=0.2895\omega_{r}$. The reason is that the greater the number of the qubits $N$, the greater the steps of the generation of GHZ state. And the fidelity error decreases with the value of ratio $g/\omega_r$ in Fig. \ref{fig_04}(c). The nonlinearity energy gap that leads to selective resonance transition depends on the coupling strength $g$. The greater the coupling strength, the better the effect of blocking transition.
However, it can be found that there are several special peaks in the Fig. \ref{fig_04}(c), which do not meet our above-mentioned characteristics. Here taking $g = 0.2263\omega_r$, $\varepsilon \approx 0.282$ for $N = 5$ as an example, the fidelity is much smaller than the ideal value. Therefore, we plot the populations of state for $N=5$ and $g = 0.2263\omega_r$ in Fig. \ref{fig_04}(d). One can find that dynamic evolution is ideal during time $[0, \tau_4]$. But the blocking transition is not perfectly realized during time $[\tau_4, \tau_5]$. After adjusting the drive frequency $\omega_d = \Delta_{k_0=4}\approx 1.6352\omega_r$ (i.e. $l_0=-1$) appropriately, selective resonance transitions $\ket{W_5^4} \leftrightarrow \ket{W_5^5}$ and $\ket{W_5^0} \leftrightarrow \ket{W_5^1}$ occur simultaneously, because the driving frequency coincidentally satisfies the relationship $2\omega_d \approx \Delta_{k_0=0}$ (i.e. $l_0=-2$) which is the resonance condition between the states $\ket{W_5^0}$ and $\ket{W_5^1}$. 
In Fig. \ref{fig_03}(a3)-(d3) and Fig. \ref{fig_04}, the reasons for other unusual peaks are similar. 
\begin{table}[h]
\centering
\caption{The creation of four-qubit GHZ state with the initial state $\ket{\psi(0)}=\ket{W_{4}^{0}}\otimes\ket{0}$, other parameters are chosen as $\epsilon=0.01\omega_{r}$, $\Delta=2\pi \times 5.4$ GHz, $\eta_{d}=0.92$, and $\omega_{r}=2\pi \times 2.2$ GHz.}
\label{tab-I}
\begin{tabular}{cccc}
\hline
\hline
  Step & $\omega_{d}$/$\omega_{r}$ & Time $t$  & Quantum states $\ket{\psi(t)}$\\
\hline
   1 & 3.1457 & $t_1\approx 1.0957\times 10^{-8}$s & $ \frac{1}{\sqrt{2}}\left(\ket{W_{4}^{0}} + i\ket{W_{4}^{1}}\right)\otimes\ket{0} $\\
  2 & 2.6849 & $t_2 \approx 2.8850 \times 10^{-8}$s & $ \frac{1}{\sqrt{2}} \left(\ket{W_{4}^{0}}- \ket{W_{4}^{2}}\right)\otimes\ket{0} $\\
  3 & 2.2241 & $t_3 \approx 4.6743\times 10^{-8}$s & $ \frac{1}{\sqrt{2}} \left(\ket{W_{4}^{0}}- i\ket{W_{4}^{3}}\right)\otimes\ket{0} $\\
  4 & 1.7633 & $t_4 \approx 6.8657 \times 10^{-8}$s & $ \frac{1}{\sqrt{2}} \left(\ket{W_{4}^{0}}+ \ket{W_{4}^{4}} \right)\otimes\ket{0}$\\
\hline
\hline
\end{tabular}
\end{table}
\begin{table}[h]
\centering
\caption{The fidelities of GHZ states from the initial excitation number $k_{0}=0$ with $g=0.2895\omega_{r}$ for different $N$.}
\label{tab-II}
\begin{tabular}{cccccc}
\hline
\hline
  GHZ states & $\ket{GHZ}_{3}$ & $\ket{GHZ}_{4}$ & $\ket{GHZ}_{5}$ & $\ket{GHZ}_{6}$ & $\ket{GHZ}_{7}$\\
\hline
   Fidelity & 0.9996 & 0.9987 & 0.9976 &0.9935 & 0.9811\\
\hline
\hline
\end{tabular}
\end{table}

\section{The thermodynamic limit case: $N\gg 1$}
\label{SecIV}

In the previous content, we have discussed and proved the important role of nonlinear energy levels in the preparation of entangled states. In this section, we will briefly discuss the effect of nonlinear energy levels in the thermodynamic limit.

In Ref. \cite{PRL2003Emary,PRE2021Gonzalez}, the operator $\hat{J}^{2}=\hat{J}_{x}^{2}+\hat{J}_{y}^{2}+\hat{J}_{z}^{2}$ can be viewed as total pseudospin operator and we can consider the maximum pseudospin $j=N/2$. This has the effect of treating the collection of $N$ two-level atoms as a single ($N+1$)-level system with pseudospin $j=N/2$. We use the Holstein-Primakoff transformation
\begin{subequations}\label{4.1}
  \begin{align}
     \hat{J}_z & = \hat{b}^\dag \hat{b}-j, \\
       \hat{J}_+ & = \hat{b}^\dag (2j-\hat{b}^\dag \hat{b})^{\frac{1}{2}},\\
       \hat{J}_- & = (2j-\hat{b}^\dag \hat{b})^{\frac{1}{2}} \hat{b}.
     \end{align}
\end{subequations}
Here $\hat{b}^{\dagger}(\hat{b})$ denote the creation (annihilation) ``magnon" operators, which satisfy the commutation relation $[\hat{b},\hat{b}^{\dag}]=1$. Taking the thermodynamic limit ($j \to \infty$) or low-lying excitations with $\langle \hat{b}^{\dagger} \hat{b}\rangle\ll 2j$. Therefore we can safely approximate $\hat{J}_{+} \approx \hat{b}^\dag \sqrt{2j}$ and $\hat{J}_{-} \approx \hat{b} \sqrt{2j}$, the Hamiltonian in Eq. (\ref{eq_03}) can be recast
\begin{equation}\label{eq_19}
     \hat{H}^{(b)}(t) = \hat{H}_{r}^{(b)} + \hat{H}_{q}^{(b)}(t) + \hat{H}^{(b)}_I,
\end{equation}
where
\begin{subequations}\label{eq_19}
  \begin{align}
     \hat{H}_{r}^{(b)}&=\omega_{r} \hat{a}^{\dagger} \hat{a},\nonumber\\
     \hat{H}_{q}^{(b)}(t) &= \left[\Delta + 2\Omega_{d} \cos(\omega_{d}t)-\frac{8g^2j}{\omega_{r}} \right] \hat{b}^\dag \hat{b}-\frac{4g^2}{\omega_{r}} (\hat{b}^\dag \hat{b})^2 , \nonumber\\
      \hat{H}^{(b)}_I &= \epsilon_{j} \left[ \hat{D} (\alpha)\hat{b}^\dag + \hat{D} (-\alpha)\hat{b} \nonumber\right].
     \end{align}
\end{subequations}
Here $\epsilon_{j}=\frac{\epsilon}{2}\sqrt{2j} $ and constant term is ignored. The Kerr type nonlinearity is induced by coupling of two-level systems and the resonator, and its strength depends on coupling constant $g$. To show the nonlinearity clearly, we assume $\Omega_{d}=0$.  It is easy
to find that $\ket{m}_{2} \otimes \ket{n}_{1}$ (where $\ket{m}_{2}$ and $\ket{n}_{1}$ are Fock states of ``magnon" and resonator, respectively) is the eigenstate of $\hat{H}_{0}^{(b)}(t) = \hat{H}_{r}^{(b)} + \hat{H}_{q}^{(b)}(t)$ with energy
\begin{equation}
  E_{n,m}=\Delta\left(m-j\right)-\frac{4g^{2}}{\omega_{r}}\left(m-j\right)^{2}+n\omega_{r}.
\end{equation}
Obviously, energy gap $\Delta_{m} =E_{n,m+1}-E_{n,m}$ between two adjacent levels is a function of excitation number $m$,
\begin{equation}\label{d}
  \Delta_{m}= \Delta + \frac{4 g^2}{\omega_r}(2j-2m-1).
\end{equation}
An appropriate drive field ($\Omega_{d}\neq0$) also can be used to engineer selective interactions. Moving to an interaction picture defined by
\begin{equation}\label{u0}
  \hat{U}_{0}^{(b)}(t) = \exp \left[ -i \int_{0}^{t} dt^{\prime} \hat{H}_{0}^{(b)}(t^{\prime}) \right],
\end{equation}
so that we can eliminate the Kerr-like term, the transformed Hamiltonian is
\begin{equation}\label{eq_20}
   \begin{split}
      \hat{\tilde{H}}_{I}^{(b)}(t) =& \hat{U}_{0}^{(b)\dagger}(t)\left[ \hat{H}^{(b)}(t)-i\partial_{t} \right]\hat{U}_{0}^{(b)}(t), \\
        =& \epsilon_{j} \hat{D} (\alpha e^{i\omega_r t}) \hat{b}^\dag e^{i\left[\Delta+\frac{4g^2}{\omega_r}(2j-2\hat{b}^{\dagger} \hat{b}-\hat{I})\right]t} e^{i\theta(t)} + \rm H.c..
   \end{split}
\end{equation}
Here $\theta(t)=\eta_{d}\sin({\omega_{d}t})$. 
With the Jacobi-Anger identity [cf. Eq. (\ref{J})], in the eigenstates basis, the Hamiltonian (\ref{eq_20}) can be recast
\begin{equation*}\label{eq_22}
  \begin{split}
      \hat{H}^{(1,2)}(t) =& \sum_{mnl} \sum_{s=1}^{\infty} \left[ \Omega^{(s)}_{jmnl}(\alpha,\eta_d) e^{i\delta^{(+ s)}_{ml} t} \hat{A}_{n+s,n} \right.\\
      &+ (-1)^s  \Omega^{(s)}_{jmnl}(\alpha,\eta_d) e^{i\delta^{(- s)}_{ml} t} \hat{A}_{n,n+s} \\
      &+ \left.\Omega^{(0)}_{jmnl}(\alpha,\eta_d) e^{i\delta^{(0)}_{ml} t} \hat{A}_{n,n}\right]\otimes \hat{B}_{m+1,m}  +\rm H.c.,
  \end{split}
\end{equation*}
where $\hat{B}_{m,m^{\prime}} = \ket{m}_2\bra{m^{\prime}}$, $\hat{A}_{n,n^{\prime}} = \ket{n}_1\bra{n^{\prime}}$ and
\begin{subequations}\label{23}
  \begin{align}
     \Omega^{(s)}_{jmnl} & = \epsilon_j \mathcal{J}_{l}(\eta_d) \sqrt { \frac{n!(m+1)}{(n+s)!}} e^{ - \frac{1}{2}\alpha^{2}}\alpha^{s} L_{n}^{(s)} (\alpha^{ 2 }),\nonumber\\
      \delta^{(\pm s)}_{ml} & = l\omega_d +\Delta_{m}\pm s\omega_{r}.\nonumber
     \end{align}
\end{subequations}
The resonance frequencies $\delta^{(\pm s)}_{ml}$ depend on $m$, $s$ and $l$. Therefore, we can tune the driving frequency $\omega_{d}$ to select different type interactions. Then the effective Hamiltonian will be obtained. It is similar to Sec. \ref{SecII}.

When we tune the driving frequency $\omega_d = -\Delta_{m_0}/l_{0}$, only the operator combinations of the kind $\hat{B}_{m_0+1,m_0} \otimes \hat{A}_{n_0,n_0}$ is time-independent. One also can verify that the following RWA conditions $\left| \delta^{(0)}_{m l} \right| \gg \left|\Omega^{(0)}_{jmnl}(\alpha,\eta_d) \right|$ for $m \neq m_0$ or $l \neq l_0$, $\left|\delta^{(+s)}_{ml} \right| \gg \left|\Omega^{(s)}_{jmnl}(\alpha,\eta_d) \right|$ and $\left| \delta^{(-s)}_{ml} \right| \gg \left|\Omega^{(s)}_{jmnl}(\alpha,\eta_d) \right|$ are satisfied in the ultrastrong coupling regime. Let the initial state be $\ket{m_0}_{2} \otimes \ket{n_0}_{1}$. Only the transition $\ket{m_0}_{2} \otimes \ket{n_0}_{1} \leftrightarrow \ket{m_0+1}_{2} \otimes \ket{n_0}_{1}$ is permitted and other transitions are suppressed. The system evolution is dominated by the following effective Hamiltonian
\begin{equation}\label{eq_27}
\small
    \hat{H}_{\rm eff}^{\rm car} = \Omega^{(0)}_{jm_0n_0l_0}(\alpha,\eta_d)  \left( \hat{B}_{m_0+1,m_0} \otimes \hat{A}_{n_0,n_0} + \rm H.c.\right).
\end{equation}

So far, the model Hamiltonian of thermodynamic limit can be simplified to the controlled selective interaction under frequency modulation. It can be found that the effective Hamiltonian is similar to the case of finite qubits [cf. Eq. (\ref{eq_12})]. Therefore, the applications of selective interactions in Sec. {\ref{SecIII}} can also be realized under the thermodynamic limit, such as, ``magnon" Fock states population trapping, the preparations of ``magnon" Fock states and so on.

\begin{figure}
  \centering
  \includegraphics[angle=0,width=0.48\textwidth]{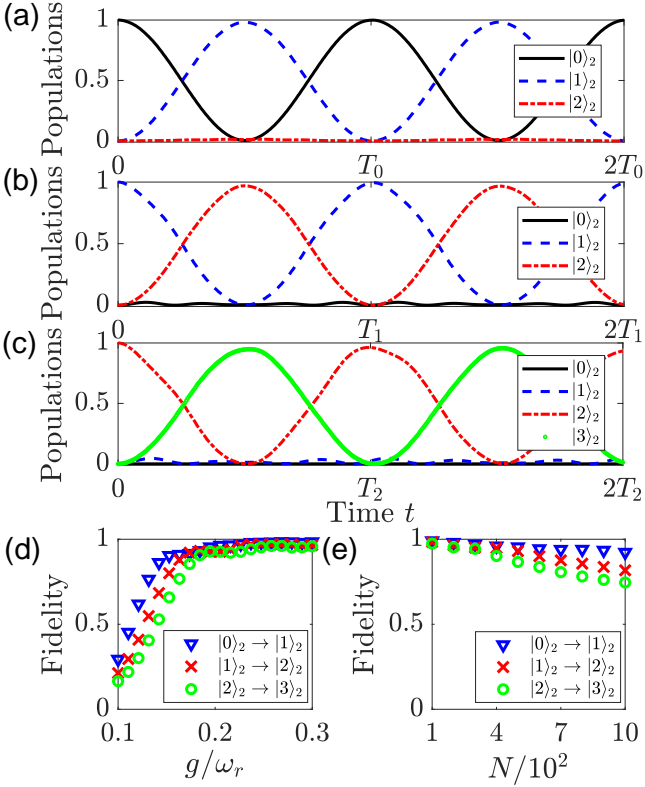}
  \caption{Numerical results of $N=200$ based on the selective resonant interaction, where the parameters are chosen as $\epsilon=0.01\omega_{r}$, $\Delta=2\pi \times 5.4$ GHz, and $\omega_{r}=2\pi \times 2.2$ GHz. (a)-(c) Populations of the states $\ket{0}_2$, $\ket{1}_2$, $\ket{2}_2$ and $\ket{3}_2$ versus time $t$ with the coupling strength $g=0.24\omega_{r}$. Time $t$ display as multiples of $T_0=\pi/[\Omega_{100,0,0,1}^{(0)}]$, $T_1=\pi/[\Omega_{100,1,0,1}^{(0)}]$, and $T_2=\pi/[\Omega_{100,2,0,1}^{(0)}]$. (d) Fidelities of the states $\ket{1}_2$, $\ket{2}_2$ and $\ket{3}_2$ generated from $\ket{0}_2$, $\ket{1}_2$ and $\ket{2}_2$ versus $g/\omega_r$, respectively. (e) Fidelities of the states versus $N$ with $g = 0.2895\omega_{r}$.}\label{fig_06}
\end{figure}

In the following content, we will show the numerical simulation of the Dicke model with driven biased term under the finite $N=200$ to verify the feasibility of the effective Hamiltonian of the selective interaction.

Given selective resonant interaction, the ``magnon" Fock states can be created easily. Taking the state $\ket{\psi_{f}}=\ket{1}_2\otimes \ket{0}_{1}$ as an example, it can be generated with a step of evolution. Initially we prepare the state $\ket{\psi(0)}=\ket{0}_2\otimes \ket{0}_{1}$ (i.e. $m_0=0$ and $n_0=0$) and apply the interaction in Eq. (\ref{eq_27}) with the driving parameters $\omega_{d} = \Delta_{m_0=0}\approx 48.3042\omega_{r}$ (i.e. $l_0=-1$) and $\Omega_{d} = 0.92\omega_{d}$ for a time period of $T_0=\pi/[\Omega_{100,0,0,1}^{(0)}]\approx 3.0091 \times 10^{-9}$s. If the large detuning constraint is fulfilled perfectly as shown in Fig. \ref{fig_06}(a), after the evolution we obtain the target state $\ket{\psi(T_0/2)}=\ket{1}_2\otimes \ket{0}_{1}$ with the fidelity $\mathcal{F}=0.9816$. In Fig. \ref{fig_06}(b) and (c), other ``magnon" Fock states $\ket{2}_2$ and $\ket{3}_2$ also can be generated with high fidelity, when the initial state is given.
It is clearly shown in Fig. \ref{fig_06} (d) that the fidelities of the states $\ket{1}_2$, $\ket{2}_2$ and $\ket{3}_2$ increases with the ratio $g/\omega_{r}$ with the following parameters $N=200$, $\epsilon=0.01\omega_{r}$, $\Delta=2\pi \times 5.4$ GHz, and $\omega_{r}=2\pi \times 2.2$ GHz. The reason is consistent with the non-thermodynamic limit. Our method relies on the energy level nonlinearity related to the coupling strength $g$.
Therefore, in order to prepare the higher fidelity quantum state, it is an effective method to appropriately increase the value of the coupling strength. In Fig. \ref{fig_06} (e), the fidelities decreases with the number of qubits.

Similarly, one can also realize the preparation of the superposition Fock states and other entangled states of ``magnon" and resonator, and the quantum states population trapping by selecting appropriate driving parameters, which will not be elaborated in this work.

\section{Conclusion}
\label{SecV}

To summarize, quantum entangled states such as Dicke states and superposition Dicke states have a wide range of applications in quantum information and quantum computing. The preparations of entangled states have always been a topic of interest. This article studies the design and preparations of high-fidelity quantum entangled states in the Dicke model with driven biased term based on the frequency modulation theory. One can verify the effectiveness of the method through numerical simulations of specific examples.

Various studies have proved that the Dicke model can be implemented in a variety of quantum systems such as circuit QED systems and hybrid magnetic cavity systems. Therefore, the realization of the driven Biased Dicke model is universal. On the other hand, the model has a wide range of applications in finite qubits and the thermodynamic limit systems.

The system can almost be ``full controllable" for the preparation of quantum states by time-periodic driving. For a set system, it can realize various entangled states such as Dicke states, superposition Dicke states, and so on by adjusting different driving parameters. Numerical simulation results show that the prepared quantum states have very high fidelity. At the same time, the ``fully controllable" of preparing a quantum state is also reflected in that one can coherently control the opening and closing of the dynamic evolution at any time by changing the driving amplitude or frequency. The population trapping of quantum states is an effective means to temporarily store the quantum states.

\section*{Acknowledgments}
The work is supported by the Fundamental Research Funds for the Central Universities (Grant Nos. 2412020FZ026 and 2412019FZ040) and Natural Science Foundation of Jilin Province (Grant No. JJKH20190279KJ).

\begin{appendix}

\section{The displacement operator in the Fock states basis}\label{appendix1}
If $\hat{A}$ and $\hat{B}$ are two noncommuting operators that satisfy the conditions $[[\hat{A},\hat{B}],\hat{A}]=[[\hat{A},\hat{B}],\hat{B}]=0$, then
\begin{equation}
   e^{\hat{A}+\hat{B}}  = e^{\hat{A}} e^{\hat{B}} e^{- \frac{1}{2} [\hat{A},\hat{B}]}.
\end{equation}
Therefore, the displacement operator $\hat{D}(\alpha e^{i\omega_r t})$ can be expanded as
\begin{equation*}
    \begin{split}
        \hat{D}(\alpha e^{i\omega_r t}) &= e^{\alpha e^{i\omega_r t} \hat{a}^{\dagger}} e^{-\alpha e^{-i\omega_r t}\hat{a}} e^{- \frac{1}{2}\alpha^2}\\
        &= e^{- \frac{1}{2}\alpha^2} \sum_{p=0}^{\infty} \frac{(\alpha e^{i\omega_r t}\hat{a}^{\dagger})^p}{p!} \sum_{q=0}^{\infty} \frac{(-\alpha e^{-i\omega_r t}\hat{a})^q}{q!}\\
        &= e^{- \frac{1}{2}\alpha^2} \sum_{p,q=0}^{\infty} \alpha^{p} (-\alpha)^{q} \frac{(\hat{a}^{\dagger})^p \hat{a}^q}{p! q!} e^{i(p-q)\omega_r t}.
    \end{split}
\end{equation*}
In the Fock state basis, the displacement operator can be reduced to
\begin{equation}\label{D}
        \hat{D}(\alpha e^{i\omega_r t}) = \sum_{m,n=0}^{\infty} \left\langle m \right| \hat{D}(\alpha e^{i\omega_r t}) \left| n \right\rangle \hat{A}_{m,n},
\end{equation}
where $\hat{A}_{m,n} = \ket{m} \bra{n}$. By using the relations
\begin{equation*}
  \hat{a}^{\dag}\ket{n}=\sqrt{n+1}\ket{n+1},\quad \hat{a}\ket{n}=\sqrt{n}\ket{n-1},
\end{equation*}
the matrix elements $D_{mn}=\left\langle m \right| \hat{D}(\alpha e^{i\omega_r t}) \left| n \right\rangle$ can be written as follows
\begin{equation}
\small
    \begin{split}
        & D_{mn}\\
        =& e^{- \frac{1}{2}\alpha^2} \sum_{p,q=0}^{\infty} \alpha^{p} (-\alpha)^{q} \frac{\left\langle m \right| (\hat{a}^{\dagger})^p \hat{a}^q \left| n \right\rangle}{p! q!}  e^{i(p-q)\omega_r t} \\
        =& e^{- \frac{1}{2}\alpha^2} \sum_{p,q=0}^{\infty} \delta_{m,n-q+p} \alpha^{p} (-\alpha)^{q} \frac{\sqrt{n!(n-q+p)!}}{p! q!(n-q)!} e^{i(p-q)\omega_r t}.
    \end{split}
\end{equation}
When $p=q$, and $m=n-q+p=n$, we can get
\begin{equation*}
    \begin{split}
        \hat{D}(\alpha e^{i\omega_r t})&= e^{- \frac{1}{2}\alpha^2} \sum_{n,q=0}^{\infty}  \frac{(-\alpha^{2})^q n!}{q! q! (n-q)!} \hat{A}_{n,n}\\
        &= e^{- \frac{1}{2}\alpha^2} \sum_{n=0}^{\infty}L_{n}^{(0)}\left(\alpha^{2}\right) \hat{A}_{n,n}.
    \end{split}
\end{equation*}
When $p=q+s$ (i.e. $p>q$), and $m=n-q+p=n+s$, we can get
\begin{eqnarray*}
   && \hat{D}(\alpha e^{i\omega_r t}) \\
   &=& e^{- \frac{1}{2}\alpha^2} \sum_{n,q=0}^{\infty} \sum_{s=1}^{\infty} e^{is\omega_r t} \alpha^{s}  \frac{(-\alpha^{2})^q \sqrt{n!(n+s)!}}{(q+s)! q! (n-q)!} \hat{A}_{n+s,n} \\
   &=& e^{- \frac{1}{2}\alpha^2} \sum_{n=0}^{\infty} \sum_{s=1}^{\infty} e^{is\omega_r t} \alpha^{s} L_{n}^{(s)}\left(\alpha^{2}\right) \sqrt{\frac{n!}{(n+s)!}}  \hat{A}_{n+s,n}.
\end{eqnarray*}
When $q=p+s$ (i.e. $p<q$), and $m=n-q+p=n-s$, we can obtain
\begin{equation*}
    \begin{split}
        &\hat{D}(\alpha e^{i\omega_r t})\\
        =& e^{- \frac{1}{2}\alpha^2} \sum_{n,p=0}^{\infty} \sum_{s=1}^{\infty} e^{-is\omega_r t} (-\alpha)^{s}  \frac{(-\alpha^{2})^p \sqrt{n!(n+s)!}}{(p+s)! p! (n-p)!} \hat{A}_{n,n+s}\\
        =& e^{- \frac{1}{2}\alpha^2} \sum_{n=0}^{\infty} \sum_{s=1}^{\infty} e^{-is\omega_r t} (-\alpha)^{s} L_{n}^{(s)}\left(\alpha^{2}\right) \sqrt{\frac{n!}{(n+s)!}}  \hat{A}_{n,n+s}.
    \end{split}
\end{equation*}
Here $L_{n}^{(s)}(\alpha^{2})$ is an associated Laguerre polynomial with $s=|m-n|$ as follows
\begin{equation}\label{2.15}
    L_{n}^{(s)}(\alpha^{2}) = \sum_{i=0}^{n} (-\alpha^{2})^{i} \frac{(n+s)!}{(s+i)! (n-i)! i!}.
\end{equation}
Then the matrix elements read
\begin{equation*}
D_{mn}=\left\{\begin{array}{ll}
e^{-\frac{1}{2}\alpha^{2}}  L_{n}^{(0)}\left(\alpha^{2}\right), & m = n; \\
e^{-\frac{1}{2}\alpha^{2}} \alpha^{s} e^{is\omega_r t}\sqrt{\frac{n !}{m !}} L_{n}^{(s)}\left(\alpha^{2}\right), & m > n; \\
e^{-\frac{1}{2}\alpha^{2}} (-\alpha)^{s}e^{-is\omega_r t} \sqrt{\frac{m !}{n !}} L_{m}^{(s)}\left(\alpha^{2}\right), & m<n.
\end{array}\right.
\end{equation*}

\end{appendix}

\bibliography{ref}

\vspace{8pt}

\end{document}